\documentclass[english,showpacs,prb,twocolumn]{revtex4}
\usepackage{graphicx}
\usepackage{enumerate}
\usepackage{hyperref}
\usepackage{textcomp}
\usepackage{amsmath}
\usepackage{amssymb}
\usepackage{pgf}

\newcommand{\revision}[1]{{\textcolor{black}{#1}}}

\newcommand{\nep}{\textrm{e}}

\newcommand{\QA}{\mathrm{\scriptscriptstyle QA}}
\newcommand{\QAIT}{\mathrm{\scriptscriptstyle QA-IT}}
\newcommand{\QART}{\mathrm{\scriptscriptstyle QA-RT}}
\newcommand{\SA}{\mathrm{\scriptscriptstyle SA}}

\newcommand{\A}{{\bf A}}
\newcommand{\B}{{\bf B}}
\newcommand{\G}{{\bf G}}
\newcommand{\F}{{\bf F}}

\newcommand{\opc}[1]{{\hat{c}^{\phantom \dagger}}_{#1}}
\newcommand{\opcdag}[1]{{\hat{c}^{\dagger}}_{#1}}
\newcommand{\opbfc}[1]{{\hat{{\bf c}}^{\phantom \dagger}}_{#1}}
\newcommand{\opbfcdag}[1]{{\hat{{\bf c}}^{\dagger}}_{#1}}

\newcommand{\PauliSigma}{\hat{\sigma}}
\newcommand{\sigmabar}{\overline{\sigma}}

\newcommand{\calN}{\mathcal{N}}

\newcommand{\Z}{{\bf Z}}

\begin{document}


\title{Quantum annealing speedup over simulated annealing on random Ising chains}

\author{Tommaso Zanca$^{1}$ and Giuseppe E. Santoro$^{1,2,3}$}

\affiliation{
$^1$ SISSA, Via Bonomea 265, I-34136 Trieste, Italy\\
$^2$ CNR-IOM Democritos National Simulation Center, Via Bonomea 265, I-34136 Trieste, Italy\\
$^3$ International Centre for Theoretical Physics (ICTP), P.O.Box 586, I-34014 Trieste, Italy
}

\begin{abstract}
We show clear evidence of a quadratic speedup of a quantum annealing (QA) Schr\"odinger dynamics over a
Glauber master-equation simulated annealing (SA) for a random Ising model in one dimension,
\revision{via an equal-footing exact deterministic dynamics of the Jordan-Wigner fermionized problems.} 
\revision{This is remarkable, in view of the arguments of Katzgraber {\em et al.}, PRX {\bf 4}, 021008 (2014), 
since SA does not encounter any phase transition, while QA does.}
We also find \revision{a second remarkable result: that a ``quantum-inspired''} imaginary-time Schr\"odinger 
QA provides a further exponential speedup, i.e., an asymptotic residual error decreasing as a power-law $\tau^{-\mu}$
\revision{of the annealing time $\tau$.}  
\end{abstract}

\pacs{75.30.Kz, 73.43.Nq, 64.60.Ht, 05.70.Jk}

\maketitle

{\it Introduction.} Quantum annealing (QA) is an offspring of thermal annealing, where the time-dependent 
reduction of quantum fluctuations
is used to search for minimal energy states of complex problems.
As such, the idea is more than two decades old \cite{Finnila_CPL94,Kadowaki_PRE98,Brooke_SCI99,Santoro_SCI02},
but it has recently gained momentum from the first commercially available
quantum annealing programmable machines based on superconducting flux quantum bits \cite{Harris_PRB10,Johnson_Nat11}.
Many problems remain open both on fundamental issues \cite{Santoro_JPA06,Das_RMP08,Dutta:book,Suzuki_EPJ15} 
and on the working of the quantum annealing machine \cite{Boixo_NatPhys13,Boixo_PRX14,Boixo_2015:arxiv}.
Among them, if-and-when QA would provide a definite speedup over simulated thermal annealing \cite{Kirkpatrick_SCI83} (SA), 
and more generally, what is the potential of QA as an optimization strategy 
for hard combinatorial problems \cite{Zamponi_QA:review,Katzgraber_PRX14,Knysh_2015:arxiv}. 
\revision{
QA seems to do better than SA in many problems \cite{Santoro_JPA06,Das_RMP08,Dutta:book,Suzuki_EPJ15} but cases are 
known where the opposite is true, notably Boolean Satisfiability (3-SAT) \cite{Battaglia_PRE05} and 3-spin antiferromagnets 
on regular graphs \cite{Polkovnikov_PRL15}.
Usually, the comparison is done by looking at classical Monte Carlo (MC) SA against Path-Integral MC (PIMC) QA 
\cite{Santoro_SCI02,Martonak_PRB02,Martonak_PRE04,Battaglia_PRE05,Stella_PRB06,Heim_SCI15}, 
but that raises issues related to the MC dynamics, especially in QA.
One of these issues has to do with the nature of the PIMC-QA dynamics, in principle not
directly related to the physical dynamics imposed by the Schr\"odinger equation, but nevertheless apparently
matching very well the experimental tests on the D-Wave hardware \cite{Boixo_NatPhys13}, 
and also showing a correct scaling of gaps in simple tunneling problems \cite{Troyer_2015:arxiv}. 
The second issue has to do with the correct time-continuum limit, requiring in principle a number of Trotter slices
$P\to \infty$ in PIMC-QA: this has led Ref.~\cite{Heim_SCI15} to question the superiority of PIMC-QA over SA 
on two-dimensional (2d) Ising spin glasses, first observed in Refs.~\cite{Santoro_SCI02,Martonak_PRB02}.
Katzgraber {\em et al.} \cite{Katzgraber_PRX14} have also questioned the ability of low-dimensional 
Ising glasses, notably in 2d but also on the Chimera graph, to be good benchmarks for the battle QA versus SA,
as SA would not encounter any finite temperature transition (the glass transition temperature is $T_g=0$),
while QA goes through a quantum phase transition \cite{Sachdev:book} in all cases.  
}


 
\revision{
Given the highly unsettled situation, any non-trivial problem in which a careful comparison between a classical SA dynamics 
and genuine QA is possible would be highly valuable: there are not many problems, beyond Grover's search \cite{Grover_PRL97,Roland_PRA02}, 
in which a clear quantum speedup is ascertain beyond doubt.  Also, ideally the comparison should be performed by relying
on a deterministic approach, avoiding issues related to the MC dynamics: that, however, usually limits us to studying
problems with just a few Ising spins, $N\sim 20$, hence not really conclusive about the regime $N\to\infty$. 
}

\revision{In this Letter we present our results} on a problem --- the one dimensional (1d) random ferromagnetic Ising model ---
where a deterministic approach can be used to compare on equal footing SA to a Schr\"odinger equation QA. 
\revision{
While the problem has no frustration, hence in some sense {\em simple} from the point of view of combinatorial
optimization\cite{Zamponi_QA:review} --- the two classical ground states are trivial ferromagnetic states with all spins aligned ---
it has nevertheless a non-trivial annealing {\em dynamics}, where disorder plays a crucial role. 
Remarkably, it is a problem which does not fit into Ref.~\cite{Katzgraber_PRX14} prescription,
since the classical annealing encounters no phase-transition at any finite temperature, while QA does encounter a quantum phase-transition:
and yet, as we will show, a definite {\em quadratic quantum speedup} can be demonstrated.
}
\revision{Technically,} for SA we will resort to studying a Glauber-type master equation with a ``heat-bath''  choice for the transition matrix, 
which allows for a Jordan-Wigner fermionization \cite{Nishimori_PRE15} of the corresponding imaginary-time quantum problem
\cite{vanKampen:book,Somma_PRL07,Boixo_EPJ15}.
For QA, the quantum fluctuations are provided by the usual transverse-field term, which is
annealed to zero during the QA evolution. 
Results for real-time Schr\"odinger QA are known for the ordered \cite{dziarmaga05,Zurek_PRL05} 
and disordered \cite{Dziarmaga_PRB06,Caneva_PRB07} Ising chain, already demonstrating 
the crucial role played by disorder, in absence of frustration:
the Kibble-Zurek \cite{kibble80,zurek85} scaling $1/\sqrt{\tau}$ of the density of defects $\rho_{\rm def}$ generated by the annealing
of the ordered Ising chain \cite{zurek96,Polkovnikov_RMP11,Dutta:book} --- $\tau$ being the total annealing time --- 
turning into a $\rho_{\rm def}\sim \log^{-2}{(\gamma \tau)}$ for the real-time Schr\"odinger QA with disorder \cite{Dziarmaga_PRB06,Caneva_PRB07}. 
We will show here that similar quality deterministic results for SA yields $\rho_{\rm def}\sim \log^{-1}{(\gamma_{\SA}\tau)}$, 
\revision{thus providing the desired evidence of a QA quadratic speedup}. 
Moreover, we will be able to compare an imaginary-time Schr\"odinger QA to the (physical) real-time QA. 
The usual conjecture is that the two approaches should have a similar asymptotic behavior \cite{Stella_PRB05,Morita_JMP08}. 
We show here that this is not true for Ising chains 
in the thermodynamic limit: 
imaginary-time QA gives $\rho_{\rm def}\sim\tau^{-2}$ for the ordered Ising chain, 
and $\rho_{\rm def}\sim \tau^{-\mu}$ with $\mu\sim 1\div 2$ in the disordered case --- 
an {\em exponential speedup}. 
\revision{This remarkable result suggests that ``quantum inspired''
algorithms based on imaginary-time Schr\"odinger QA might be a valuable root in quantum optimization.}

{\it Model and methods.} The problem we deal with is that of classical Ising spins, $\sigma_j=\pm 1$, 
in \revision{1d} with nearest-neighbor ferromagnetic random couplings $J_j>0$, 
$H_{\rm cl}= - \sum_{j=1}^L J_j \sigma_j \sigma_{j+1}$. 
We study its SA classical annealing dynamics, as described by a Glauber master equation (ME) \cite{Glauber_JMP63}
\begin{equation} \label{Glauber_ME:eqn}
\frac{\partial P(\sigma,t)}{\partial t} = 
\sum_j W_{\sigma, {\sigmabar}^j} P({\sigmabar}^j,t) - \sum_j W_{{\sigmabar}^j, \sigma} P(\sigma,t) \;.
\end{equation}
Here $\sigma=(\sigma_1,\cdots,\sigma_L)$ denotes a configuration of all $L$ spins,  
with a probability $P(\sigma,t)$ at time $t$, 
${\overline{\sigma}}^j = (\sigma_1,\cdots,-\sigma_j,\cdots,\sigma_L)$ is a
configuration with a single spin-flip at site $j$, 
and $W_{{\sigmabar}^j,\sigma}$ is the transition matrix from $\sigma$ to $\sigmabar^j$. 
$W$ will depend on the temperature $T$, which is in turn decreased as a function of time, $T(t)$, 
to perform a ``thermal annealing''.
Many different choices of $W$ are possible, all satisfying the detailed balance (DB) condition
$W_{\sigma,\sigma'} \, P_{\rm eq}(\sigma') = W_{\sigma',\sigma} \,P_{\rm eq}(\sigma)$,
where $P_{\rm eq}(\sigma)=\nep^{- \beta H_{\rm cl}(\sigma)}/Z$ is the Gibbs distribution at fixed $\beta=1/(k_BT)$ 
and $Z$ the canonical partition function. 
For all these choices of $W$, the Glauber ME can be turned into
an imaginary-time (IT) Schr\"odinger problem \cite{vanKampen:book,Somma_PRL07,Nishimori_PRE15} 
by ``symmetrizing $W$'' into an Hermitean ``kinetic energy'' operator $K$, with the help of DB and the substitution
$P(\sigma,t) = \sqrt{ P_{\rm eq}(\sigma) } \; \psi(\sigma,t)$. 
\revision{
This technique was exploited in Ref.~\cite{Kaneco_JPSJ15} to re-derive Geman\&Geman bound \cite{Geman_IEEE84} on the SA optimal schedule.
}
\revision{Here, it} leads to 
$-{\partial_t \psi(\sigma,t)} = - \sum_j K_{\sigma,\sigmabar^j} \, \psi(\sigmabar^j,t) + V(\sigma) \, \psi(\sigma,t)$
with $K_{\sigmabar^j,\sigma} = K_{\sigma,\sigmabar^j} = W_{ \sigmabar^j,\sigma} \; \sqrt{ P_{\rm eq}(\sigma)/P_{\rm eq}(\sigmabar^j)}$ 
and $V(\sigma) = \sum_j W_{\sigmabar^j,\sigma}$. 
The crucial step forward comes from the discovery of Ref.~\onlinecite{Nishimori_PRE15} that the {\em heat-bath} choice of 
$W_{\sigmabar^j,\sigma}=\alpha \; \nep^{-\beta H_{\rm cl}(\sigmabar^j)}/(\nep^{-\beta H_{\rm cl}(\sigma)}+\nep^{-\beta H_{\rm cl}(\sigmabar^j)})$,
$\alpha$ being an arbitrary rate constant, leads to a Schr\"odinger problem which is quadratic when expressed in terms 
of Jordan-Wigner fermions.  
In operator form, we can write our heat-bath SA problem as an IT Schr\"odinger equation:\cite{nota:betadot}
%
%
\begin{equation} \label{SA:eqn}
-\frac{\partial}{\partial t} |\psi(t)\rangle = \widehat{H}_{\SA}(t) | \psi(t) \rangle \;. 
\end{equation}
The ``quantum'' Hamiltonian $\widehat{H}_{\SA} = - \widehat{K}_{\SA} + \widehat{V}_{\SA}$,
can be readily expressed in terms of Pauli matrices: 
$\widehat{K}_{\SA}  = \sum_j \Gamma_{j}^{(0)} \PauliSigma^x_j - 
\sum_j \Gamma_{j}^{(2)} \PauliSigma^z_{j-1} \PauliSigma^x_j \PauliSigma^z_{j+1}$ and
$\widehat{V}_{\SA} =  -\sum_j \Gamma_{j}^{(1)} \PauliSigma^z_j \PauliSigma^z_{j+1} + \frac{\alpha}{2} L$,
where the couplings $\Gamma_{j}^{(0,1,2)}$ have simple expressions in terms of $\cosh(\beta J_{j})$ and $\sinh(\beta J_{j})$
\cite{Nishimori_PRE15,Zanca:unpub}.
A Jordan-Wigner transformation \cite{Lieb_AP61} makes $\widehat{H}_{\SA}$ 
quadratic in the fermionic operators $\opc{j}$ and $\opcdag{j}$ \cite{Nishimori_PRE15,nota:JW}.

Consider now the quantum annealing (QA) approach to the same problem. 
For that, one would add to $H_{\rm cl}$ a transverse field term whose coupling $\Gamma(t)$ is slowly turned off,  
obtaining a transverse field random Ising model \cite{Fisher_PRB95}
$\widehat{H}_{\QA}(t) = -\sum_j J_j \PauliSigma^z_j \PauliSigma^z_{j+1} - \Gamma(t) \sum_j \PauliSigma^x_j$,
with a Schr\"odinger dynamics governed by 
\begin{equation} \label{QA:eqn}
\xi \frac{\partial}{\partial t} |\psi(t)\rangle = \widehat{H}_{\QA}(t) | \psi(t) \rangle \;. 
\end{equation}
Here $\xi=i$  (with $\hbar=1$) for the (physical) real-time (RT) dynamics --- we dub it QA-RT ---, 
while $\xi=-1$ for an IT dynamics --- QA-IT in short. 

In all cases, both SA and QA, the resulting Hamiltonian can be cast into a quadratic BCS fermionic form: 
\begin{multline} \label{quadratic-H:eqn}
\widehat{H}(t) = \left( \begin{array}{cc}  \opbfcdag{} & \opbfc{} \end{array} \right)
  \left( \begin{array}{cc} {\phantom{+}\A}(t) & \phantom{+}{\B}(t) \\
                          -{\B}(t) & -{\A}(t) \end{array} \right)
     \left( \begin{array}{l}  \opbfc{} \\ \opbfcdag{} \end{array} \right) \;,
\end{multline}
where the $2L\times 2L$ matrix formed by $L\times L$ blocks $\A$ (symmetric) and $\B$ (antisymmetric),
couples the fermionic operators, 
$(\opcdag{1} \; \cdots \opcdag{L} \; \opc{1} \cdots \opc{L}) = (\opbfcdag{} \; \opbfc{}\!\!)$.
The form of $\A$ and $\B$ is given in Ref.~\onlinecite{Caneva_PRB07} for QA, in 
Refs.~\onlinecite{Nishimori_PRE15,Zanca:unpub} for SA.

The most efficient way to solve \eqref{QA:eqn}-\eqref{quadratic-H:eqn} for $\xi=i$ 
is through the Bogoliubov-de Gennes (BdG) equations \cite{dziarmaga05,Dziarmaga_PRB06,Caneva_PRB07}. 
%
In IT, this approach leads to an {\em unstable} algorithm, due to
\revision{the difficulty of maintaining orthonormality for a set of $L$ exponentially blowing/decaying BdG solutions}.
To do IT dynamics, \revision{to solve Eq.~\eqref{SA:eqn} and Eq.~\eqref{QA:eqn} with $\xi=-1$, we introduce a different strategy.} 
%
The most general BCS state has a Gaussian form \cite{Ring:book}:
\begin{equation}
|\psi(t) \rangle = 
{\calN}(t) \; \exp{\Big(\frac{1}{2} \displaystyle\sum_{j_1j_2} \Z_{j_1j_2}(t) \opcdag{j_1} \opcdag{j_2}\Big) } \; |0\rangle \;,
\end{equation}
where ${\calN}(t)$ is a normalization constant, and $\Z$ \revision{an $L\times L$ antisymmetric matrix}. 
%
As a quadratic $\widehat{H}(t)$ conserves the Gaussian form of $|\psi(t)\rangle$ \cite{Ring:book}, 
one can transform \cite{Zanca:unpub} 
\eqref{SA:eqn} or \eqref{QA:eqn} into a first-order non-linear differential equation for $\Z$:
\begin{equation} \label{Zevolution:eqn}
\xi \, \dot{\Z} = 2 \Big( \A \cdot \Z + \Z \cdot \A + \B + \Z \cdot \B \cdot \Z \Big) \;.
\end{equation}
%
%
All physical observables can be calculated from Wick's theorem \cite{Ring:book}, once the Green's functions 
$\G_{j'j} (t)= \langle \psi(t) | \opcdag{j} \opc{j'} | \psi(t) \rangle$ and
$\F_{j'j} (t) = \langle \psi(t) | \opc{j} \opc{j'} | \psi(t) \rangle$ are known. 
Simple algebra \cite{Zanca:unpub} shows that $\G = ({\bf 1} + \Z \Z^{\dagger})^{-1} \Z\Z^{\dagger}$
and $\F = ({\bf 1} + \Z \Z^{\dagger})^{-1} \Z$.
The defects acquired over the classical ferromagnetic ground state (GS) with all spin aligned are
antiparallel pairs of spins, \revision{measured} by $(1-\PauliSigma^z_j\PauliSigma^z_{j+1})/2$,
whose average follows from $\langle\psi(t)| \PauliSigma^z_j\PauliSigma^z_{j+1} |\psi(t) \rangle = (\G_{j+1,j} + \F_{j,j+1}+ c.c.)$. 

{\em Results.}
To monitor the annealing, we calculate the density of defects over the ferromagnetic classical GS  
$\rho_{\rm def}(t) =\sum_{j} \langle\psi(t)| (1-\PauliSigma^z_j\PauliSigma^z_{j+1})|\psi(t)\rangle/(2L)$
%
%
and the residual energy density 
$\epsilon_{\rm res}(t) = \sum_{j} J_j \, \langle\psi(t)| (1- \PauliSigma^z_j\PauliSigma^z_{j+1}) |\psi(t)\rangle/L$
%
%
for SA, Eq. \eqref{SA:eqn}, QA-RT and QA-IT, Eq.~\eqref{QA:eqn}. 
For simplicity we considered a linear decrease of the annealing parameter, with a total annealing time
$\tau$: for SA we set $T(t)=T_0(1-t/\tau)$, for QA   
$\Gamma(t)=\Gamma_0(1-t/\tau)$, with $t$ from $0$ to $\tau$. 
The initial temperature ($T_0$) or transverse field ($\Gamma_0$)
are set to reasonably large values, $k_BT_0=5\div 10$ and $\Gamma_0=5\div 10$, 
both in units of the $J$-coupling.

\begin{figure}[ht]
\begin{center}
  \includegraphics[width=8.0cm]{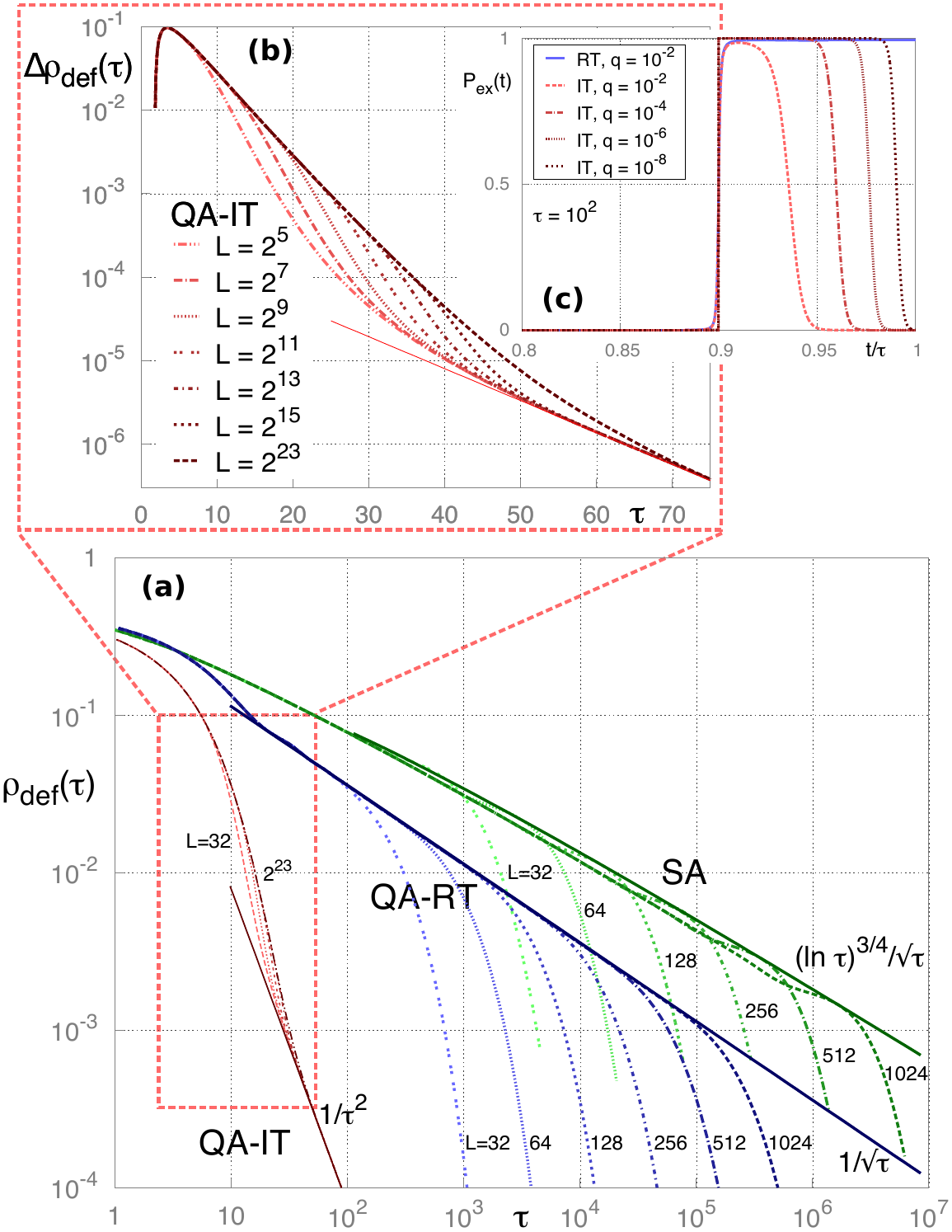}
\end{center}
\caption{(Color online) 
Density of defects after the annealing, $\rho_{\rm def}(\tau)$, versus the annealing time $\tau$ 
for the ordered Ising chain. 
(a) Results for simulated annealing (SA), for quantum annealing (QA) in real time (QA-RT) and in imaginary time (QA-IT). 
(b) Log-linear plot of the deviation $\Delta \rho_{\rm def}=\rho_{\rm def}-a/\tau^2$, with $a\approx 0.784$,
for QA-IT, showing a clear exponential approach to the leading $1/\tau^2$.
(c) Landau-Zener dynamics in real and imaginary time for modes close to the critical wave-vector, $k=\pi-q$
with small $q$. }
\label{Ordered:fig}
\end{figure}

We start from the simpler problem of the ordered Ising chain.
The previous general approach simplifies when all $J_j=J$ and periodic boundary conditions (BC) are used:
the Hamiltonian reduces to a collection of $2\times 2$ problems, 
\begin{multline} 
\widehat{H}(t) = \sum_{k>0} \left( \begin{array}{cc}  \opcdag{k} & \opc{-k} \end{array} \right)
  \left( \begin{array}{cc} a_k(t) & \phantom{+} b_k(t) \\
                                     b_k(t) & -a_k(t) \end{array} \right)
     \left( \begin{array}{l}  \opc{k} \\ \opcdag{-k} \end{array} \right) \;, \nonumber
\end{multline}
%
%
in terms of $k$-space fermions $(\opcdag{k},\opc{-k})$. 
The BCS state is  
$|\psi(t)\rangle=\prod_{k>0} {(1+|\lambda_k(t)|^2)}^{-1/2} \, \nep^{-\lambda_k(t) c_k^\dagger c_{-k}^\dagger}|0\rangle$,
the Schr\"odinger dynamics in Eq.~\eqref{Zevolution:eqn} reduces to \cite{nota:ordered}
$\xi \, \dot{\lambda}_k=2\lambda_k(t) \, a_k(t)-b_k(t) +\lambda_k^2(t) \, b_k(t)$.
The behavior of the final density of defects $\rho_{\rm def}(t=\tau)$ for real-time QA follows the
Kibble-Zurek power-law $\rho_{\rm def}^{\QART}(\tau) \sim 1/\sqrt{\tau}$ associated to crossing the 
Ising critical point \cite{dziarmaga05,Zurek_PRL05}, see Fig.~\ref{Ordered:fig}(a).
Finite-size deviations, revealed by an exponential drop of $\rho_{\rm def}(\tau)$, 
occur for annealing times $\tau\propto L^2$, 
due to a Landau-Zener (LZ) probability of excitation across a small gap $\Delta_k\sim \sin{k} \sim 1/L$
close to the critical wave-vector $k_c=\pi$, 
$P_{\rm ex}=\nep^{-\alpha \Delta_k^2 \tau}$. 
We note that, for finite $L$, the exponential drop of $\rho_{\rm def}^{\QART}(\tau)$ eventually turns
into a $1/\tau^2$, due to finite-time corrections to LZ \cite{Vitanov_PRA99,Caneva_PRB08}.
%
%
The QA-IT case is very different from QA-RT for $L\to \infty$. 
We find $\rho_{\rm def}^{\QAIT}(\tau) \sim a/\tau^2 + O(\nep^{-b\tau})$, 
%
%
where the first term is due to non-critical modes, while the exponentially decreasing term 
(see Fig.~\ref{Ordered:fig}(b)) is due to critical modes with $k=\pi-q$ at small $q$:
their LZ dynamics, see Fig.~\ref{Ordered:fig}(c), shows that IT follows a standard LZ up
to the critical point, but then {\em filters the ground state} (GS) exponentially fast 
as the gap resurrects after the critical point.
That IT evolution gives different results from RT for $L\to \infty$ is not obvious. 
From the study of toy problems \cite{Stella_PRB05}, it was conjectured 
that QA-IT might have the same asymptotic behavior as QA-RT, as later shown more
generally \cite{Morita_JMP08} from adiabatic perturbation theory estimates.
That is what happens in our Ising case too for {\em finite} $L$ and $\tau\to \infty$,
with a common $1/\tau^2$ asymptotic.   
Moreover, IT gives the same critical exponents as RT for QA ending at the critical point \cite{DeGrandi_PRB11}.
The deviation of QA-IT from QA-RT for Ising chains {\em in the thermodynamic limit $L\to \infty$}
is due to the non-perturbative LZ nature when the annealing proceeds beyond the critical point.  
The SA result, Fig.~\ref{Ordered:fig}(a), is marginally worse than QA-RT due to logarithmic corrections,
$\rho_{\rm def}^{\SA}(\tau) \sim(\ln{\tau})^{\nu}/\sqrt\tau$, 
where we find \cite{nota:SA} $\nu\approx 3/4$.

\begin{figure}[ht]
\begin{center}
  \includegraphics[width=8cm]{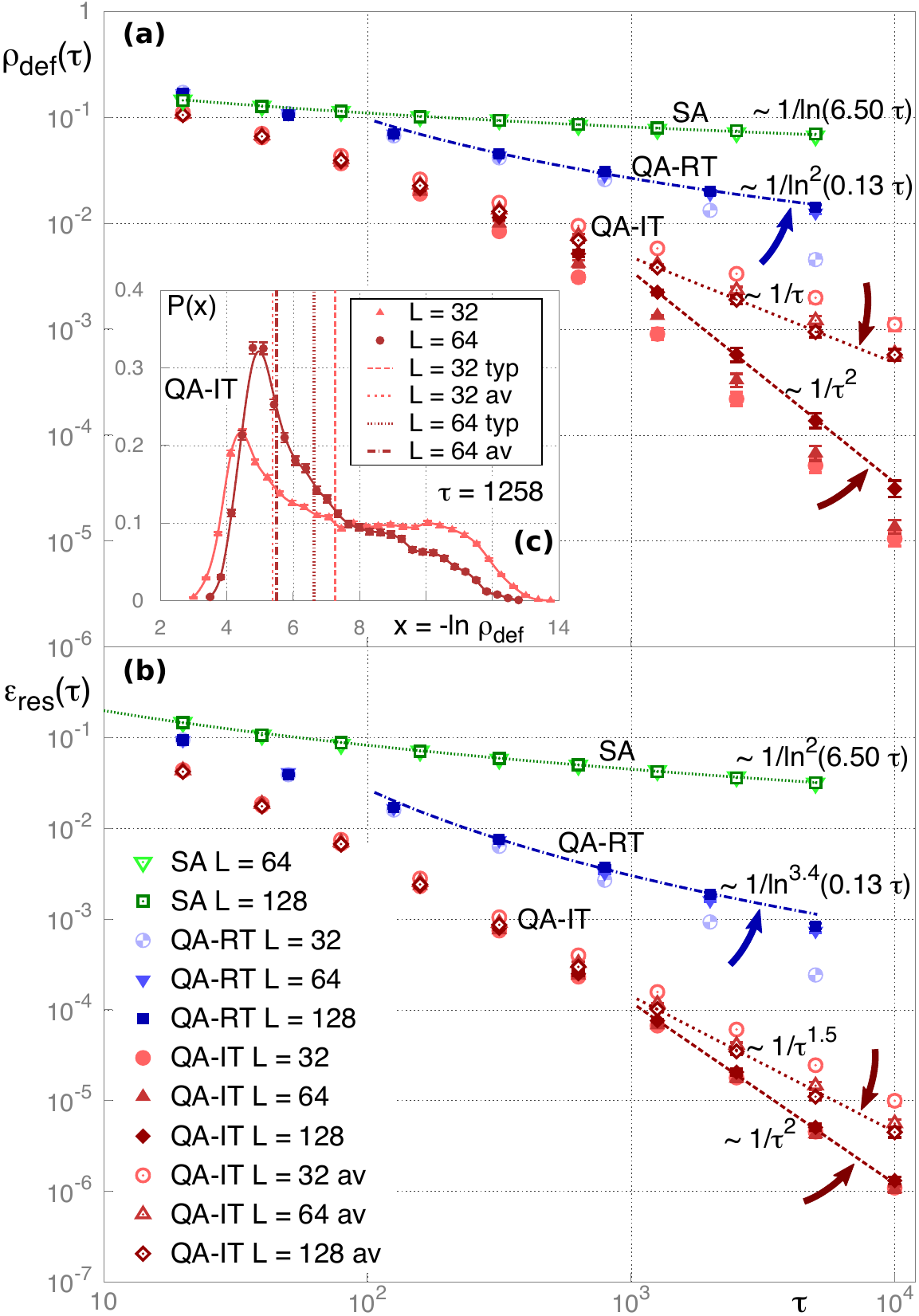}
\end{center}
\caption{(Color online) 
(a) Density of defects $\left[\rho_{\rm def}(\tau)\right]_{\rm typ}$ 
and (b) residual energy $\left[\epsilon_{\rm res}(\tau)\right]_{\rm typ}$ 
versus $\tau$ for SA, QA-RT, and QA-IT (for which average data are also shown).
The lines are fits of the data. Solid arrows are guides to the eye for the size dependence.
(c) Probability distribution of  
$x=-\ln \rho_{\rm def}$ for QA-IT with $\tau=1258$ for $L=32$ and $64$. 
Vertical lines denote average and typical values.
}
\label{Disordered:fig}
\end{figure}

We now turn to annealing results for disordered Ising chains with open BC, and 
couplings $J_j$ chosen from a flat distribution, $J_j\in [0,1]$. 
For QA, the transverse field random Ising model is known to possess an 
{\em infinite randomness critical point} \cite{Fisher_PRB95}, here at $\Gamma_c=1/e$,  
where the distribution of the equilibrium gaps $\Delta$ becomes a universal 
function \cite{Young1996} of $g=-(\ln\Delta)/\sqrt{L}$.
The SA Hamiltonian $\widehat{H}_{\SA}$ shows different physics: the smallest typical equilibrium gaps 
are seen at the end of the annealing, $T\to 0$, where they vanish Arrhenius-like, $\Delta_{\rm typ}(T)\sim \nep^{-B/T}$ with $B/J\sim 2$.
Turning to dynamics, we calculate $\rho_{\rm def}(\tau)$ and $\epsilon_{\rm res}(\tau)$ by integrating
numerically the equation for $\Z$ in Eq.~\eqref{Zevolution:eqn},
\revision{feasible for $L$ up to $O(1000)$. Given the need for a good statistics, we will present data up to $L=128$.}
For any given $\tau$, we considered many disorder realizations, obtaining distributions for
$\rho_{\rm def}(\tau)$ and $\epsilon_{\rm res}(\tau)$. 
For SA these distributions are approximately log-normal, as previously 
found for QA-RT \cite{Caneva_PRB07}, 
%
%
with a width decreasing as $1/\sqrt{L}$, 
implying that the average $\left[\rho_{\rm def}\right]_{\rm av}$
approaches the {\em typical} value $\left[\rho_{\rm def}\right]_{\rm typ}=\nep^{\left[\ln \rho_{\rm def}\right]_{\rm av}}$
for large $L$, and similarly for $\epsilon_{\rm res}$. 
QA-IT behaves differently: the distributions of both 
$\rho_{\rm def}(\tau)$ and $\epsilon_{\rm res}(\tau)$ show marked deviations from log-normal,
see Fig.~\ref{Disordered:fig}(c), hence typical and average values are rather different.
Fig.~\ref{Disordered:fig} shows $\left[\rho_{\rm def}(\tau)\right]_{\rm typ/av}$ (a) and
$\left[\epsilon_{\rm res}(\tau)\right]_{\rm typ/av}$ (b) for the three annealings performed. 
%
%
The SA results are nearly size-independent, with a clear logarithmic behaviour \cite{Suzuki_JSTAT09} for large $\tau$:
\begin{equation}
\left[\rho_{\rm def}\right]^{\SA} \sim \log^{-1}(\gamma_{\SA} \tau) \;, \hspace{2mm}
\left[\epsilon_{\rm res}\right]^{\SA} \sim \log^{-2}(\gamma_{\SA} \tau) \;,
\end{equation}
with $\gamma_{\SA}\approx 6.5$. 
Notice that $\epsilon_{\rm res} \sim \log^{-\zeta_{\SA}}(\gamma_{\SA}\tau)$ with $\zeta_{\SA}=2$
saturates the bound $\zeta_{\SA}\le 2$ for thermal annealing in glassy systems \cite{Huse_Fisher_PRL86}.
Concerning the QA-RT case, results are well established from Ref.~\onlinecite{Caneva_PRB07} where
larger systems \revision{were} tackled by the linear BdG equations: 
\begin{equation}
\left[\rho_{\rm def}\right]^{\QART}\!\! \sim \log^{-2}(\gamma\tau) \;, \hspace{2mm}
\left[\epsilon_{\rm res}\right]^{\QART}\!\! \sim \log^{-\zeta}(\gamma\tau) \;, 
\end{equation}
with $\gamma\approx 0.13$, and $\zeta\approx 3.4$. 
%
Finally, we again find QA-IT very different from QA-RT, 
with a faster, power-law, decrease of $\rho_{\rm def}$ and $\epsilon_{\rm res}$.
%
The size-dependence of the data is revealing: 
the ``typical'' data move upwards with increasing $L$, 
but, luckily, the ``average'' data show the opposite tendency --- they move towards lower values, 
with an increasing slope vs $\tau$.
It is fair to conclude that our data support a power-law for both
$\rho_{\rm def}$ and $\epsilon_{\rm res}$:  
%
\begin{equation}
\left[\rho_{\rm def}\right]_{\rm typ/av}^{\QAIT} \sim \tau^{-\mu_{\rho}} \;, \hspace{2mm} 
\left[\epsilon_{\rm res}\right]_{\rm typ/av}^{\QAIT} \sim \tau^{-\mu_{\epsilon}} \;,
\end{equation}
where we estimate $\mu_{\rho}\sim 1\div 2$ and $\mu_{\epsilon}\sim 1.5\div 2$.

{\em Discussion.}
We have presented a non-trivial example of a quantum speedup of real-time Schr\"odinger QA over master-equation
SA on an equal-footing single-flip deterministic dynamics. 
Our second important result is that a ``fictitious'' imaginary-time QA behaves very differently from 
the ``physical'' real-time QA, 
providing a much faster annealing, with an asymptotic behavior compatible with $\tau^{-\mu}$,
with $\mu\approx 1\div 2$, i.e., an {\em exponential speedup}. 
Hence, provocatively, ``quantum inspired'' is here better than ``quantum'' , a point that deserves further studies.
\revision{Results on the fully-connected Ising ferromagnet confirm that this IT-speedup is not specific to the present 1d problem\cite{Wauters_unpub}.}

The specific problem we addressed --- a random ferromagnetic Ising chain --- is ``easy'' in many
respects: {\em i)} it does not possess frustration, the ingredient that makes optimization problems generally 
hard \cite{Zamponi_QA:review}, {\em ii)} it can be reduced to a quadratic fermionic problem, and 
{\em iii)} is also a case where SA does not encounter any phase transition for $T\to 0$, 
while the QA dynamics goes through a critical point at $\Gamma_c>0$.  
This, as discussed in Ref.~\onlinecite{Katzgraber_PRX14} for the spin-glass case, might in principle give an unfair advantage to 
SA over QA: \revision{but, remarkably}, it doesn't, in the present case. 
Our study provides a useful benchmark for many possible developments, 
like the role of thermal effects, or the comparison with QA simulated by path-integral MC \cite{Troyer_2015:arxiv}. 
\revision{Our QA-IT results suggest} also to pursue the application of diffusion quantum MC to 
simulate the imaginary-time Schr\"odinger QA, likely a very good ``quantum-inspired'' classical 
optimization algorithm \cite{Pilati_2015:arxiv}.  

\acknowledgments{
GES wishes to thank M. Troyer for discussions and hospitality at ETH Z\"urich, where this work started,  
S. Knysh, H. Nishimori, R. Fazio and E. Tosatti for fruitful conversations.  
Research was supported by MIUR, through PRIN-2010LLKJBX-001, and by EU through ERC MODPHYSFRICT.}
   

\begin{thebibliography}{60}
\bibitem[{not({\natexlab{a}})}]{nota:betadot}
\bibinfo{note}{In the annealing case \cite {Nishimori_PRE15} one should add an
  extra term $- ({\protect \mathaccentV {dot}05F\beta }/2) (H_{\protect \rm cl}
  - \delimiter "426830A H_{\protect \rm cl} \delimiter "526930B _{\protect \rm
  eq})$ to the potential $V(\sigma )$, originating from $\protect \mathaccentV
  {dot}05F{P}_{\protect \rm eq}/P_{\protect \rm eq}$. As argued in Ref.\protect
  \nobreakspace {}\protect \onlinecite {Nishimori_PRE15}, this additional term
  is not important in the limit of a very large system.}
\bibitem[{not({\natexlab{b}})}]{nota:JW}
\bibinfo{note}{The relevant terms can be rewritten as $\PauliSigma^x_j = 2
  \,\opcdag{j}\opc{j} -1$, $\PauliSigma^z_{j-1} \PauliSigma^x_{j}
  \PauliSigma^z_{j+1} = - ( \opcdag{j-1}\opc{j+1} + \opcdag{j-1}\opcdag{j+1} +
  {\rm H.c.} )$, and $\PauliSigma^z_j \PauliSigma^z_{j+1} = ( \opcdag{j}
  \opc{j+1} + \opcdag{j} \opcdag{j+1} + {\rm H.c.})$.}
\bibitem[{not({\natexlab{c}})}]{nota:ordered}
\bibinfo{note}{The $k$-vectors are $k=(2n-1)\pi /L$ with $n=1\protect \cdots
  L/2$. For QA $a_k(t)=-2\left [\Gamma (t)+J\protect \qopname \relax
  o{cos}k\right ]$ and $b_k =2J\protect \qopname \relax o{sin}k$, while for SA
  $a_k(t)=-\left [\Gamma _0(t)+\Gamma _1(t)\protect \qopname \relax
  o{cos}k+\Gamma _2(t)\protect \qopname \relax o{cos}2k\right ]$, and
  $b_k(t)=\Gamma _1(t) \protect \qopname \relax o{sin}k +\Gamma _2(t) \protect
  \qopname \relax o{sin}2k$, with $\Gamma _0(t)=\alpha \protect \frac {\protect
  \qopname \relax o{cosh}^2{\beta J}}{\protect \qopname \relax o{cosh}{2\beta
  J}}$, $\Gamma _1(t)=\alpha \protect \qopname \relax o{tanh}{2\beta J}$,
  $\Gamma _2(t)=\alpha \protect \frac {\protect \qopname \relax o{sinh}^2{\beta
  J}}{\protect \qopname \relax o{cosh}{2\beta J}}$. The critical point is for
  $k\to \pi $ in all cases. The defect density is $\rho_{\rm def}(t) =
  (2/L)\sum_{k>0}
  \left|\lambda_k(t)\sin{(k/2)}-\cos{(k/2)}\right|^2/(1+\left|\lambda_k(t)\right|^2)$
  and $\epsilon_{\rm res}=2J \rho_{\rm def}$.}
\bibitem[{not({\natexlab{d}})}]{nota:SA}
\bibinfo{note}{Two aspects make the SA dynamics different from the standard LZ
  dynamics behind QA-RT, and are at the origin of the logarithmic corrections:
  first, the critical point occurs at $T=0$ (for $k=\pi $) and is never crossed
  during the annealing; second, the coefficients $a_k$ and $b_k$, which behave
  as $a_k\sim -2\alpha \protect \tmspace +\thinmuskip {.1667em}e^{-4\beta J}
  +\protect \mathcal {O}(\pi -k)^2$ and $b_k\sim 2\alpha \protect \tmspace
  +\thinmuskip {.1667em}(\pi -k)\protect \tmspace +\thinmuskip
  {.1667em}e^{-2\beta J}$ close to the critical point, lead to an energy gap
  which decreases exponentially fast for $T\to 0$.}
\end{thebibliography}

\end{document}